\documentclass[11pt]{article}
\usepackage{amsmath}
\usepackage{amssymb}
\def\be{\begin{equation}}
\def\ee{\end{equation}}
\renewcommand{\theequation}{\arabic{section}.\arabic{equation}}
\def\ba{\begin{eqnarray}}
\def\ea{\end{eqnarray}}
\def\nn{\nonumber}
\def\lb{\label}
\def\bb{\bibitem}
\def\R{{\cal R}}
\def\X{\mathbf{X}}%
\def\J{\mathbf{J}}
\def\L{\mathbf{L}}
\def\a{\boldsymbol{\alpha}}
\def\b{\boldsymbol{\beta}}
\def\c{\boldsymbol{\gamma}}
\begin{document}

\begin{titlepage}

\date{24 May 2018}

\title{
\begin{flushright}\begin{small}    LAPTH-016/18
\end{small} \end{flushright} \vspace{1cm}
Hairy black hole solutions in a \\ three-dimensional Galileon model}

\author{G\'erard Cl\'ement$^a$\thanks{Email: gclement@lapth.cnrs.fr} and
Khireddine Nouicer$^{b}$\thanks{Email: khnouicer@yahoo.fr} \\ \\
$^a$ {\small LAPTh, Universit\'e Savoie Mont Blanc, CNRS,} \\ {\small 9 chemin de Bellevue,
BP 110, F-74941 Annecy-le-Vieux cedex, France} \\
$^b$ {\small LPTh, Department of Physics, University of Jijel,} \\
{\small BP 98, Ouled Aissa, Jijel 18000, Algeria}}

\maketitle

\begin{abstract}
We investigate stationary rotationally symmetric solutions of a particular truncation of Horndeski theory in 
three dimensions, including a non-minimal scalar kinetic coupling to the curvature.
After discussing the special case of a vanishing scalar charge, which includes most of the previously known
solutions, we reduce the general case to an effective mechanical model in a three-dimensional target space.
We analyze the possible near-horizon behaviors, and conclude that black hole solutions with degenerate
horizons and constant curvature asymptotics may exist if the minimal and non-minimal scalar
coupling constants have the same sign.
In a special case, we find a new analytic rotating black hole solution with scalar hair and degenerate horizon. 
This is geodesically and causally complete, and asymptotic to the extreme BTZ metric. We also briefly discuss 
soliton solutions in another special case.
\end{abstract}
\end{titlepage}
\setcounter{page}{2}

\section{Introduction}
In recent years, with the advent of the dark energy paradigm to alleviate many puzzling problems in the
standard cosmological model, such as the late accelerating expansion of the universe and the coincidence problem,
many extensions of general relativity have been explored. Particularly a plethora of scalar-tensor models with
higher-derivative couplings but second-order field equations have been discovered many years ago by Horndeski
\cite{horndeski}, and recently revived as Galileon gravity models \cite{gal}. These theories are known to be free
from the Ostrogradski instabilities. Furthermore, black holes are generically hairy in such theories, as shown in \cite{SZ},
disproving a no-hair theorem of \cite{HN}.
A number of asymptotically flat or asymptotically AdS black hole solutions have been obtained
\cite{rinaldi,bhgal}.

While not directly relevant to real-world physics, lower dimensional gravity constitutes a powerful test
laboratory to handle theoretical situations which are more complicated in four dimensions. For instance,
the BTZ black hole solution \cite{btz} of three-dimensional gravity with a negative cosmological constant
has been used to test ideas about gravitational collapse, black hole thermodynamics and quantum gravity.
Moreover, black hole solutions to gravity in three dimensions may be uplifted to near-horizon solutions
to some five-dimensional \cite{brem5,brquad}, or in special cases four-dimensional \cite{brquad},  theories
of gravity. Black hole solutions to a three-dimensional theory of gravity with a minimally coupled scalar
field have been found in \cite{virbhadra}, and further discussed in \cite{GC99}. These are not asymptotically
flat or constant curvature. It was shown in \cite{BGH2} that the BTZ black hole metric
yields an exact solution of the field equations for a particular truncation of the Horndeski action
(given in (\ref{ac}) below). Other exact solutions of the same model, including warped AdS$_3$ black hole
solutions \cite{GC07,ALPSS}, were found in \cite{GT} for special relations between the model coupling constants.

It is the purpose of this paper to extend these results by making a systematic investigation of all the possible black hole
solutions of this three-dimensional model. The action and the field equations are given in the next section, and reduced
to those of a one-dimensional mechanical problem with four degrees of freedom under the assumption of two commuting
Killing vectors. This problem is further reduced in section 3, where it is shown that either the scalar charge vanishes,
in which case all the solutions are given, or the scalar field can be eliminated away to yield an effective mechanical
model in three-dimensional target space. The possible near-horizon behaviors are systematically analyzed in section 4, and
the possibility for these to be consistent with a constant curvature asymptotic behavior is discussed in section 5. Two
special cases are discussed more fully in section 6, where in particular a new asymptotically AdS and singularity-free black
hole solution is presented. Our results are summarized in the Conclusion. In an Appendix we briefly discuss a by-product of our
analysis concerning the existence of soliton solutions in at least one special case.

\setcounter{equation}{0}
\section{Action and field equations}

We will study the following part of the Horndeski action in three-dimensional gravity
\be\lb{ac}
S=\pm\int d^{3}x\sqrt{-g}\left[\R-2\Lambda-\frac{1}{2}\left(\alpha g^{\mu\nu}-\eta G^{\mu\nu}\right)
\partial_{\mu}\phi\partial_{\nu}\phi\right]
\ee
where $\R$ is the trace of the Ricci tensor, $G^{\mu\nu}$ is the contravariant Einstein
symetric tensor and $\phi$ is a scalar field, dubbed the Galileon. With an arbitrary overall normalization,
the model contains only three real coupling constants $\Lambda$ (cosmological constant), $\alpha$
(minimal scalar coupling constant) and $\eta$, Horndeski theory proper corresponding to $\eta\neq0$,
which we shall assume hereafter. The overall sign $\pm$ accounts for the sign of the Newton
gravitational constant, which is arbitrary in three-dimensional gravity \cite{DJH}.
Throughout this paper we use the mostly plus convention $(-++)$ for the signature of Lorentzian metrics.
The variation of the action with respect to the metric and the scalar field gives respectively
\be
G_{\mu\nu}=-\Lambda g_{\mu\nu}+\frac{\alpha}{2}\left[\partial_{\mu}\phi\partial_{\nu}\phi-\frac{1}{2}
g_{\mu\nu}\left(\partial\phi\right)^{2}\right]-\frac{\eta}{4}
\left[\left(\partial\phi\right)^{2}G_{\mu\nu}+
g_{\mu\rho}\delta_{\nu\gamma\kappa}^{\rho\sigma\tau}\nabla^{\gamma}\nabla_{\sigma}\phi
\nabla^{\kappa}\nabla_{\tau}\phi\right]
\ee
and
\be
\nabla_{\mu}\left[\left(\alpha g^{\mu\nu}-\eta G^{\mu\nu}\right)\nabla_{\nu}\phi\right]=0
\ee
where $(\partial\phi)^2 = g^{\mu\nu}\partial_\mu\phi\partial_\nu\phi$, and
$\delta_{\nu\gamma\kappa}^{\rho\sigma\tau}=\varepsilon_{\nu\gamma\kappa}\varepsilon^{\rho\sigma\tau}$,
with $\varepsilon^{\rho\sigma\tau}$ the three-dimensional antisymmetric symbol.

In this paper we follow a powerful approach to solve the equations
of motion in the stationary, rotationally symmetric case (two commuting Killing vectors).
The approach consists in reducing the problem to a mechanical
one in an abstract three-dimensional Minkowski space. The approach
developed in  \cite{GC94a,GC93} is based on the
observation that the $SL\left(2,R\right)$ group of transformations
is locally isomorphic to the Lorentz group $SO\left(2,1\right),$
suggesting that the general stationary rotationally symmetric metric
may be written in the 2+1 form
\be\lb{para}
ds^{2}=\lambda_{ab}\left(\rho\right)dx^{a}dx^{b}+\zeta^{-2}\left(\rho\right)R^{-2}\left(\rho\right)d\rho^{2},
\ee
 where $x^0=t$, $x^1=\varphi$, $\lambda$ is the $2\times2$ matrix
\be
\lambda=\left(\begin{array}{cc}
T+X & Y\\
Y & T-X
\end{array}\right),
\ee
$-\textrm{det}\lambda=R^{2}=\mathbf{X}^{2}$ is the Minkowskian
norm of the vector $\mathbf{X}=\left(T,\:X\:,Y\right)$
\be\lb{mink}
\mathbf{X}^{2}=\eta_{ij}X^iX^j=-T^{2}+X^{2}+Y^{2},
\ee
and the positive function $\zeta\left(\rho\right)$ allows for arbitrary reparameterizations
of the radial coordinate $\rho$. The stationary sector corresponds to spacelike $\X$ ($R^2>0$).
The ADM form of the metric (\ref{para}) is
 \be\lb{adm}
ds^2 = - \frac{R^2}Vdt^2 + \frac{d\rho^2}{\zeta^2R^2} + V\left(d\varphi+\frac{Y}Vdt\right)^2
 \ee
with $V = T-X$, so that zeroes of $R^2$ with positive $V$ correspond to Killing horizons.

With the parametrization (\ref{para}), the action (\ref{ac})
is reduced to the form
 \be\lb{ac2}
S=\pm\int d^{2}x\int d\rho L
 \ee
 where the generalized Lagrangian $L(\zeta,\X,\X',\phi,\phi')$ reads
\be\lb{lag}
L = -2(\zeta \X\cdot\X')' + \frac{1}{2}\zeta\mathbf{X}'^{2} - 2\zeta^{-1}\Lambda -
\frac{1}{2}\zeta\left(\alpha-\frac{\eta}{4}\zeta^{2}\mathbf{X}'^{2}\right)\X^2\phi'^{2}
\ee
(the prime indicates a derivative with respect to $\rho$, and the dot product is the Minkowskian scalar product).
The first term in the Lagrangian (\ref{lag}) is a pure divergence and may be discarded. The Euler
equation relative to the cyclic variable $\phi$ may be first integrated to
\be\lb{phi}
\left(\alpha-\frac{\eta}{4}\zeta^{2}\mathbf{X}'^{2}\right)\zeta R^{2}\phi' = C
\ee
where $C$ is an integration constant, which may be thought of as the scalar charge. Variation with respect to $\zeta$ gives the
Hamiltonian constraint
\be\lb{ham}
\left(\alpha-\frac{3\eta}{4}\zeta^2\mathbf{X}'^{2}\right)R^{2}\phi'^{2} = \mathbf{X}'^{2}+4\zeta^{-2}\Lambda.
\ee
Finally the transverse Einstein equations follow from the variation relative to $\X$:
\be\lb{ein}
\left(\zeta\Psi\mathbf{X}'\right)'=-\zeta\phi^{'2}\left(\alpha-\frac{\eta}{4}\zeta^2\mathbf{X}'^{2}\right)\mathbf{X},
\ee
where
 \be\lb{Psi}
\Psi=1+\frac{\eta}{4}\zeta^2R^{2}\phi^{'2}.
 \ee
Also, as shown in \cite{GC94b}, invariance of the action (\ref{ac2}) under $SO\left(2,1\right),$ leads to the
conservation of the generalized angular momentum $\J = \X\wedge(\partial L/\partial\X')$,
 \be\lb{J}
\mathbf{X}\wedge\left(\zeta\Psi\mathbf{X}'\right)=\mathbf{J},
 \ee
where the wedge product is defined by $({\bf X} \wedge {\bf Y})^i =$
$\eta^{ij}\epsilon_{jkl}X^k Y^l$ (with $\epsilon_{012} = +1$).
These last three equations are not independent equations, but first integrals of the system (\ref{phi})-(\ref{ein}).
At this point, we recall that the gauge funtion $\zeta(\rho)$ is an arbitrary function (invariance under general
transformations of the radial coordinate $\rho$), and choose in the following the gauge $\zeta=1$.

\setcounter{equation}{0}
\section{Reduction of the field equations}
\subsection{Case $C=0$, $\alpha\neq0$}
The corresponding solution in the static (diagonal) case was previously given in \cite{BGH2}. Here we extend it
to the stationary case. Assuming $\phi'\neq0$ (for constant $\phi$ the field equations reduce trivially
to the Einstein equations with cosmological constant), we obtain from eq. (\ref{phi})
 \be\lb{C0}
\X^{'2} = \frac{4\alpha}\eta.
 \ee
If $\alpha\neq0$, then equations (\ref{ham}) and (\ref{ein}) reduce to
\be
\phi^{'2}=\frac{2(\Lambda_0-\Lambda)}{\alpha R^{2}},\quad
\Psi=\frac{\Lambda_0+\Lambda}{2\Lambda_0},\quad\Psi\mathbf{X}''=0, \quad \X^{'2} = -4\Lambda_0,
\ee
where we have put $\Lambda_0 \equiv -\alpha/\eta$.

If \underline{$\alpha\neq\Lambda\eta$} ($\Lambda_0+\Lambda\neq0$), the last two equations are equivalent to
the stationary rotationally symmetric vacuum Einstein equations (eqs. (\ref{ein}) and (\ref{ham})
with $\alpha=\eta=0$) for an effective cosmological constant $\Lambda_0$, with for instance BTZ
black hole solutions for $\Lambda_0<0$ ($\eta/\alpha>0$) \cite{BGH2}.

If \underline{$\alpha=\Lambda\eta$} ($\Lambda=-\Lambda_0$), then $\Psi=0$ so that the transverse Einstein equations
(\ref{ein}) are identically satisfied. The three metric functions $\X$ are to a large extent arbitrary,
subject to the sole scalar constraint (\ref{C0}). The scalar field $\phi$ is then obtained by integrating
 \be
\phi^{'2}=-\frac4{\eta R^{2}},
 \ee
and is real provided $\eta<0$. This is the {\em degenerate} sector discussed in \cite{BGH2} in the case of
static solutions. We extend here this discussion to the more general case of stationary solutions. A
generic stationary metric can be parameterized in the form (\ref{para}), which may be written for $\zeta=1$ as
\be
ds^2 = Udt^2 + 2Ydtd\varphi + Vd\varphi^2 + \frac{d\rho^2}{Y^2-UV},
\ee
where $U=T+X$, $V=T-X$, $Y$ are three independent functions of $\rho$. In the case $C=0$,
$\alpha=\Lambda\eta\neq0$, they are subject to the sole constraint (\ref{C0}), which reads
 \be\lb{C01}
Y^{'2} - U'V' = \frac{4\alpha}\eta,
 \ee
leaving us with two arbitrary independent functions plus one integration constant. In the static case,
$Y(\rho)=0$, there remains only one arbitrary function, e.g. $U(\rho)$, the function $V$ (noted $r^2$ in
\cite{BGH2}) being related to $U$ by $U'V'=-4\alpha/\eta$, which is equivalent to eq. (7) of \cite{BGH2}.
If for instance we choose instead  in the stationary case $V(\rho)=0$, the function $U(\rho)$ is again arbitrary,
while the function $Y(\rho)$ is given (after a suitable translation of the radial coordinate $\rho$) by
$Y=c\rho$, with $c^2=4\alpha/\eta$ (provided $\alpha/\eta>0$), corresponding to the metric
 \be
ds^2 = U(\rho)dt^2 + 2c\rho\,dt\,d\varphi + \frac{d\rho^2}{c^2\rho^2},
 \ee
which after setting $\rho\propto r^{-2}$ is the AdS-wave solution (4) of \cite{GT} (discussed in the
degenerate case in the sentence after (12) of the same paper). The dual solution ($U(\rho)=0$)
 \be
ds^2 = - \frac{c^2\rho^2}{V(\rho)}dt^2 + \frac{d\rho^2}{c^2\rho^2} + V(\rho)\left(d\varphi + \frac{c\rho}{V(\rho)}dt\right)^2
 \ee
corresponds to a black hole provided the arbitrary function $V(\rho)$ is analytic in $\rho$ and $V(0)>0$.
The warped AdS space and black hole
solutions (for which $\X^{'2}$ is known to be constant \cite{ACGL}) also discussed in \cite{GT} exist only
under the constraint $\alpha=\Lambda\eta$ (which translates into $\zeta\eta=\Lambda\beta$ in
the notations of \cite{GT}, see their eq. (20)), so that they are also a very special case of these
degenerate solutions.

\subsection{Case $C=0$, $\alpha=0$}
This case deserves a special treatment. Eq. (\ref{C0}) with $\alpha=0$ is
 \be\lb{C00}
\X^{'2} = 0.
 \ee
The left-hand side of (\ref{ham}) then vanishes, while the right-hand side vanishes only if \underline{$\Lambda=0$}.
These two equations are then identically satisfied, so that the function $\phi(\rho)$ is left arbitrary.
Accordingly, the function $\Psi(\rho)$ defined by (\ref{Psi}) is also arbitrary. The right-hand side
of (\ref{ein}) vanishes, so that this equation can be first integrated to
 \be\lb{Xlin}
\X = \boldsymbol{\alpha}\sigma + \boldsymbol{\beta},
 \ee
where $d\sigma = \Psi^{-1}(\rho)\,d\rho$, and $\boldsymbol{\alpha}$, $\boldsymbol{\beta}$ are two arbitrary vectors with $\boldsymbol{\alpha}^2=0$ from (\ref{C00}).

In the special case $\Psi(\rho)=1$, the equations (\ref{C00}) and (\ref{Xlin}) reduce to those of vacuum Einstein gravity
(with $\Lambda_0=0$). So in the general case the spacetime metric is simply obtained from that of vacuum gravity by multiplying the metric function
$g_{\rho\rho}$ by $\Psi^2$. As discussed in \cite{GC94a}, there are four cases:

a) $\boldsymbol{\alpha}$ past lightlike: a representative solution derived from the particle-like metric
\cite{DJH} is
 \be
ds^2 = - (dt-\omega d\varphi)^2 + a^2r^2d\varphi^2 + \Psi^2(r)dr^2
 \ee
($\omega$ and $a$ constant). This will be particle-like (only delta-function singularities) if $\Psi(0)$ is finite.

b) $\boldsymbol{\alpha}$ future lightlike: a representative Rindler-like solution \cite {GC85} is
 \be
ds^2 = -r^2\left(1+\frac{\omega^2r^2}{\nu^2}\right)dt^2 + \nu^2\left(d\varphi+\frac{\omega r^2}{\nu^2}dt\right)^2 + \Psi^2(r)dr^2
 \ee
($\omega$ and $\nu$ constant).

c) $\boldsymbol{\alpha}\cdot\boldsymbol{\beta}=0$ (where $\a\cdot\b$ is the scalar product with the Minkowskian metric (\ref{mink})):
a representative solution \cite {GC85} is
 \be
ds^2 = - r(dt-\omega d\varphi)^2 + 2c(dt-\omega d\varphi)d\varphi + \Psi^2(r)dr^2
 \ee
($\omega$ and $c$ constant). This metric is such that $R^2=c^2$, and so is horizonless.

d) $\boldsymbol{\alpha}=0$: a representative solution is
 \be
ds^2 = -dt^2 + \nu^2d\varphi^2 + \Psi^2(r)dr^2.
 \ee

\subsection{Case $C\neq0$}
The scalar field $\phi(\rho)$ (or more correctly, its first derivative) may be eliminated between equations
(\ref{phi})-(\ref{ein}) to yield a system of differential equations and constraints involving only the metric
functions $\X(\rho)$. Equivalently, this elimination may be achieved at the Lagrangian level by the Routhian
method \cite{goldstein}. After discarding the pure divergence contribution to (\ref{lag}), the reduced Lagrangian is defined by
 \be
L_{\rm red}(\zeta,\X,\X') = L - \Pi_\phi\phi',
 \ee
where $\Pi_\phi \equiv (\alpha-\eta\zeta^{2}\mathbf{X}'^{2}/4)\zeta R^{2}\phi'$. After inverting (\ref{phi}) to obtain $\phi'(\zeta,\X,\X')$,
we obtain the reduced Lagrangian
 \be\lb{L1}
L_{\rm red}=\frac{1}{2}\zeta\mathbf{X}'^{2}-2\zeta^{-1}\Lambda+\frac{\zeta^{-1}C^{2}}{2\X^{2}\left(\alpha-\eta\zeta^2\mathbf{X}'^{2}/4\right)}.
 \ee
Varying with respect to $\zeta$, and then gauge-fixing $\zeta=1$, we obtain the Hamiltonian constraint
 \be\lb{ham1}
\left(4\Lambda+\mathbf{X}'^{2}\right)\left(\alpha - \frac{\eta}{4}\mathbf{X}'^{2}\right)^{2}-\left(\alpha-\frac{3\eta}{4}\mathbf{X}'^{2}\right)\frac{C^{2}}{R^{2}} = 0,
 \ee
which is a first integral of the equations obtained by varying with respect to $\X$,
 \be\lb{ein1}
\left(\Psi\mathbf{X}'\right)' = \frac\Omega{R^4}\X,
 \ee
where
 \be\lb{psiom}
\Psi = 1 + \frac{\eta C^2}{4R^2(\alpha-\eta\X'^2/4)^2},
\qquad \Omega = -\frac{C^2}{\alpha-\eta\X'^2/4},
 \ee
three other first integrals being the generalized angular momentum equations (\ref{J}).
If $\X'^2\neq4\alpha/3\eta$, the expression of $\Psi$ can be simplified using (\ref{ham1}) to
 \be\lb{psi1}
\Psi=\frac{\alpha+\eta\Lambda-\eta\X'^2/2}{\alpha-3\eta\X'^2/4}
 \ee
(the special case $\X'^2=4\alpha/3\eta$ shall be discussed further in Sect. 5).

These equations are similar to the equations for the motion of a particle in a central potential $W(R)$,
which here is defined implicitly by inverting (\ref{ham1}) to rewrite it in the form
 \be\lb{energy}
\frac12\X'^2 + W(R) = 0.
 \ee
A noteworthy difference, however, is that in the present case $\X'^2$ is not positive definite. As in the classical central potential problem, the present one
may be reduced to a one-dimensional ``radial'' problem by squaring (\ref{J}) and eliminating $\X'^2$ from (\ref{energy}) to yield
 \be\lb{radial}
R'^2 - \frac{J^2\Psi^{-2}(R)}{R^2} + 2W(R) = 0,
 \ee
where $J^2 = \J^2$ is a real constant (note the minus sign in (\ref{radial}), due to the indefinite target
space metric). In the following two sections we analyze the information on possible black-hole solutions which
may be obtained from these equations, before presenting in section 6 several explicit solutions in
special cases.

\setcounter{equation}{0}
\section{Near-horizon behavior}
The master equation (\ref{ham1}) is a cubic equation in the variable
$\mathbf{X}'^{2}$. It can be put in normal form
 \be\lb{cubic}
y^3+Py-Q = 0,
 \ee
with
 \be\lb{yPQ}
y = \X'^2 + \frac{4(\Lambda+2\Lambda_0)}3, \quad  P=48\Lambda^2\left(Z-k^2\right),\quad
Q=128\Lambda^3\left[\left(1-\frac{3k}{2}\right)Z-k^3\right],
 \ee
and
 \be
k = \frac{\Lambda-\Lambda_0}{3\Lambda}, \quad Z = \frac{C^2}{4\eta\Lambda^2R^2}
 \ee
($\Lambda_0=-\alpha/\eta$).
The number of real solutions to the cubic equation (\ref{cubic}) depends on the sign of the discriminant
$\Delta=\left(P/3\right)^3+\left(Q/2\right)^2$ which is given by
\be\lb{Discriminant}
\Delta=(4\Lambda)^6Z\left[Z^2+\left(1-3k-\frac{3}{4}k^2\right)Z+2k^3\left(3k-1\right)\right]
\ee
The sign of the discriminant depends on the sign of $Z$ (the sign of $\eta$) and on the sign of the binomial in $Z$ between the square brackets. The sign of the latter depends in turn on the sign of its discriminant, which reads
\be\lb{Discriminant-prime}
\Delta'=\frac{1}{16}\left(2+3k\right)\left(2-5k\right)^3.
\ee
It follows that, for $\eta>0$, $\Delta$ is negative, and our master equation (\ref{cubic}) has three real roots, if
$-2/3 < k < 2/5$ ($-3 < \alpha/\eta\Lambda < 1/5$) and $Z$ lies between the two roots of the binomial in
(\ref{Discriminant}). For $\eta<0$, the situation is opposite, so that under the same conditions $\Delta$ is positive
and our master equation has only one real root.

We can draw an immediate consequence concerning black holes. The Killing horizon corresponds to $R=0$, i.e.
to $Z \to \pm\infty$, according to the sign of $\eta$. So for $\eta>0$ there will be at most one black
hole solution (provided $V$ is positive for $R^2=0$), while for $\eta<0$ three black hole
solutions could in principle coexist for the same
values of the model coupling constants and of the scalar charge $C$.

Now we investigate what are the possible near-horizon behaviors allowed by the reduced field equations.
From our master equation (\ref{ham1}), $R^2 \to 0$ is possible for $C\neq0$ only if either a) $\X'^2
\to 4\alpha/3\eta$, or b) $\X'^2\to\infty$, leading to $(\X'^2)^2\simeq-12C^2/\eta R^2$, which is possible
only for $\eta<0$. We shall consider successively these two possibilities.

\subsection{$\X'^2 \to 4\alpha/3\eta$}
We here assume $\alpha\neq0$ (the case $\alpha=0$ is discussed in Sect. 5). In this case the function $\Psi$
defined by (\ref{psiom}) diverges. The small $R$ behaviors of the functions occurring in the
reduced field equations are
 \be\lb{41}
W(R) \simeq -\frac{2\alpha}{3\eta}, \quad \Psi(R) \simeq \frac{9\eta C^2}{16\alpha^2R^2}, \quad \Omega(R) \simeq -\frac{3C^2}{2\alpha}.
 \ee
Inserting these into the radial equation (\ref {radial}) leads to
 \be\lb{rad41}
R'^2\simeq J^2\beta^2R^2 + \frac{4\alpha}{3\eta},
 \ee
with $\beta=\alpha^2/9\eta C^2$. $R(\rho)$ can go to zero only if \underline{$\alpha/\eta>0$}. Then (\ref{rad41}) can be integrated
near $R=0$ to
 \be
R \simeq a\rho \qquad (a^2 = 4\alpha/3\eta).
 \ee
The vector field equation (\ref{ein1}) then reads
 \be
(\rho^{-2}\X')' + 2\rho^{-4}\X \simeq 0,
 \ee
which is solved by
 \be
\X = \a\rho+\b\rho^2+O(\rho^3)
 \ee
with $\a$ and $\b$ two constant target space vectors. To lowest order, this leads to the correct behaviors
of $R^2$ and $X'^2$ provided $\a^2=a^2$. From (\ref{J}) the constant angular momentum vector is
 \be
\J = \frac{C^2}{a^6\eta}\a\wedge\b.
 \ee
To complete the analysis, we recover the near-horizon behavior of the scalar field from (\ref{phi}),
 \be
\phi' \simeq \frac{9C\eta}{8\alpha^2\rho^2}.
 \ee
A typical possible near-horizon behavior, obtained by transforming the spacelike vector $\a$ to $\a = (0,-a,0)$,
and the radial coordinate $\rho$ to $r= (a\rho)^{1/2}$, is thus:
 \be\lb{nh}
ds^2 \simeq - r^2dt^2 + \frac{3\eta}\alpha\frac{dr^2}{r^2} + r^2d\varphi^2,
\qquad \phi \simeq - \frac{3C}{2\alpha ar^2}.
 \ee
Note that the horizon is always degenerate in this case.

\subsection{$\X'^2 \to \infty$}
In this case, which necessitates $\eta<0$, the behaviors are
 \be
W(R) \simeq -\frac{\gamma}{R}, \quad \Psi(R) \simeq \frac{2}{3} - \frac{2(\alpha+3\eta\Lambda)}
{9\eta\gamma}R, \quad \Omega(R) \simeq -\frac{2\gamma R}{3},
 \ee
with $\gamma^2 = -3C^2/\eta$, $\gamma$ real.

If $J^2\neq0$, the effective potential in the radial equation
(\ref{radial}) is dominated by the centrifugal term behaving as $-9J^2/4R^2$, so that there is no horizon
for $J^2<0$. For $J^2>0$, eq. (\ref{radial}) can be integrated to $R^2 \simeq 3J\rho$. Eq. (\ref{ein1})
then reads
 \be
\left(\Psi\mathbf{X}'\right)' \simeq -\frac{2\gamma}{3(3J\rho)^{3/2}}\X,
 \ee
Choosing the spacelike vector $\J$ orthogonal to the plane $(T,X)$ ($Y=0$), this equation has a solution
of the form
 \be\lb{Winf1}
U \simeq -a\rho + b\rho^{3/2}, \quad V \simeq \frac{3J}a + c\rho^{1/2}.
 \ee

If $J^2=0$, then
 \be
R'^2 \simeq \frac{2\gamma}R
 \ee
(where the positive root of $\gamma^2$ must be chosen), which integrates to
 \be
R \simeq a\rho^{2/3} \quad \left(a^3 = \frac{9\gamma}2\right).
 \ee
The approximate vector field equation (\ref{ein1})
 \be
\X'' + \frac2{9\rho^2}\X \simeq 0
 \ee
is then solved by
 \be
\X \simeq \a\rho^{1/3} + \b\rho^{2/3}.
 \ee
The behavior $\X^2 \simeq a^2\rho^{4/3}$ is recovered from this provided
 \be\lb{a0bs}
\a^2 = \a\cdot\b = 0, \quad \b^2 = a^2,
 \ee
the null vector $\J$ being
 \be
\J = \frac29\a\wedge\b = \varepsilon\frac{2a}9\a
 \ee
($\varepsilon^2=1$). Transforming to the radial coordinate $r = a\rho^{1/3}$, and choosing for simplicity the basis vectors
$\a=(1/2a,-1/2a,0)$, $\b=(0,0,a)$, we obtain the near-horizon metric
 \be\lb{Winf2}
ds^2 \simeq -r^3dt^2 + 9a^{-4}dr^2 + a^{-2}r\left(d\varphi+ardt\right)^2.
 \ee

A straightforward geodesic analysis shows however that in both cases the Killing horizon $r=0$ is not traversable,
so that the corresponding implicit solutions are not black holes. This conclusion could also have been
reached by computing the Ricci scalar \cite{GC09}:
 \be
\R = \frac12\X'^2 - (R^2)'' = 3W - 2\X\cdot\X'' \simeq 3(W-\Omega R^{-2}) \simeq - \frac\gamma{R}
 \ee
(where we have used (\ref{ein1}) with $\Psi\simeq2/3$), which diverges on the Killing horizon $R=0$. So, while
there can be in principle three different solutions with horizon for each set of values of the model coupling
constants and of the scalar charge, only one of these (that with $\X'^2 \to 4\alpha/3\eta$ on the horizon)
can correspond to a black hole.

\setcounter{equation}{0}
\section{Asymptotic behavior}
We now investigate whether black hole solutions (which are possible only for $\alpha/\eta>0$
according to the results of the previous section) can be asymptotically particle-like, i.e. asymptotic to
a constant curvature spacetime (de Sitter, Minkowski, or anti-de Sitter). As discussed in \cite{GC94a},
this corresponds to the boundary condition
 \be\lb{aspart}
\X'(\infty) = \a,
 \ee
with $\a$ a constant timelike, null or spacelike vector. This is integrated by
 \be\lb{Xinf}
\X \sim \a\rho + \b,
 \ee
leading to $R^2 \propto \rho^2$ if $\a^2\neq0$ or $R^2 \propto \rho$ if $\a$ is null. In both cases
$R^2$ goes to infinity with $\rho$, so that the master equation (\ref{ham1}) leads to
 \be\lb{haminf}
(4\Lambda+\a^2)\left(\alpha - \frac{\eta}{4}\a^2\right)^2 = 0,
 \ee
so that either $\a^2=-4\Lambda$, or $\a^2=4\alpha/\eta$. We first assume $\Lambda+\alpha/\eta\neq0$.

If \underline{$\a^2=-4\Lambda$}, equation (\ref{psiom}) shows that $\Psi$ and $\Omega$ are asymptotically
constant, so that the second order differential equation (\ref{ein1}) leads to $\X'' \sim 0$, consistent
with (\ref{aspart}). We shall present in the next section an exact black hole solution with this
asymptotic behavior. This is a solution for the special relation between coupling constants
$\Lambda+\alpha/3\eta=0$ and so is asymptotically AdS ($\Lambda<0$). There seems to be no obstruction
for an asymptotically flat black hole solution to exist when $\Lambda=0$.

If instead \underline{$\a^2=4\alpha/\eta$} ($\a$ spacelike), we find $\Psi(\infty)= (\alpha-\eta\Lambda)/2\alpha$ and
$\Omega \propto R$. For $\alpha\neq\eta\Lambda$, (\ref{ein1}) is again asymptotically $\X'' \sim 0$,
consistent with (\ref{aspart}). For $\alpha=\eta\Lambda$ (implying $\Lambda>0$),
the discussion is more subtle. Eq. (\ref{ham1}) leads to the asymptotic relation
 \be
\left(\X'^2-\frac{4\alpha}\eta\right)^2 \sim \frac{\lambda^2}{R^2},
 \ee
with $\lambda^2=-4C^2/\eta$ (provided $\eta<0$). This in turn leads to
 \be
\Psi \sim \frac{\eta\lambda}{4\alpha R} \sim \frac\lambda{a^3\rho}, \quad \Omega \sim -\lambda R
\sim -\lambda a\rho
 \ee
($a^2=4\alpha/\eta$) and to the asymptotic behavior of the differential equation (\ref{ham1})
 \be
\left(\rho^{-1}\X'\right)' + \rho^{-3}\X \sim 0,
 \ee
which is solved by $\X \sim \a\rho + \c\rho\ln\rho$, consistent with (\ref{aspart}) only if the leading
behavior is suppressed, $\c=0$, and $\b=0$ in (\ref{Xinf}). The metric is in this case asymptotic
to the BTZ vacuum.

In the special case \underline{$\Lambda+\alpha/\eta=0$}, the master equation (\ref{ham1}) leads to the
asymptotic behavior
 \be\lb{specialas}
\X'^2-\frac{4\alpha}\eta \sim \mathrm{O}(R^{-2/3}).
 \ee
Accordingly,
 \be
\Psi \sim 1, \quad \Omega \sim  \mathrm{O}(R^{2/3}),
 \ee
with $R \sim a\rho$ ($a^2=4\alpha/\eta$). Inserting this in (\ref{ham1}) leads to the asymptotic
differential equation $\X'' \sim \mathrm{O}(\rho^{-10/3})\X$. The resulting asymptotic behavior with
the leading term $\a\rho$,
 \be
\X \sim \a\left[\rho +  \mathrm{O}(\rho^{-1/3})\right],
 \ee
is not consistent with (\ref{specialas}). So we conclude that in this special case there is no
asymptotically particle-like solution.

\setcounter{equation}{0}
\section{Special solutions}
The cubic equation (\ref{ham1}) factorizes simply in two cases: $\Lambda+\alpha/3\eta=0$, and
$\alpha=0$. These two intersect in a third special case $\Lambda=\alpha=0$.

\subsection{Case $\Lambda+\alpha/3\eta=0$}
Equation (\ref{ham1}) reduces to
 \be
(\X'^2+4\Lambda)\left[(\X'^2+12\Lambda)^2 + \frac{12C^2}{\eta R^2}\right] = 0.
 \ee
There are three obvious solutions to this equation: \\

\noindent a) \underline{$\X'^2+4\Lambda=0$} \\

Then, assuming $\Lambda\neq0$,
 \be
W(R) = 2\Lambda, \quad \Psi(R) = 1 + \frac{C^2}{16\eta\Lambda^2 R^2},
\quad \Omega(R) = \frac{C^2}{2\eta\Lambda},
 \ee
and the radial equation reads
 \be
R'^2 - \frac{J^2R^2}{(R^2+C^2/16\eta\Lambda^2)^2} + 4\Lambda = 0.
 \ee
We shall discuss here only the special subcase $J^2=0$, for which this equation reduces to
 \be
R'^2 = \frac4{l^2}
 \ee
(provided $\Lambda=-l^{-2}<0$), which is solved (up to a translation of the radial coordinate $\rho$) by
 \be\lb{R21}
R^2 = \frac{4\rho^2}{l^2}.
 \ee
The resulting second order differential equation (\ref{ein1}), which for $\eta>0$ takes the form
 \be
\left[\left(1+\frac{\lambda^2}{\rho^2}\right)\X'\right]'= -\frac{2\lambda^2}{\rho^4}\X
\quad \left(\lambda = \frac{Cl^3}{8\eta^{1/2}}\right),
 \ee
is integrated by
 \be\lb{XaFbrho}
\X = \a\,F(\rho) + \b\,\rho,
 \ee
with $\a$ and $\b$ two constant vectors, and
 \be\lb{Feta+}
F(\rho) = \frac{\rho}{\lambda}\arctan\frac{\lambda}{\rho}.
 \ee
For $\eta<0$ we define $\mu^2 = -C^2l^6/64\eta a^6$ and similarly obtain the solution (\ref{XaFbrho}), with now
 \ba\lb{Feta-}
F(\rho) &=& \frac{\rho}{\mu}{\rm artanh}\frac{\mu}{\rho} \quad (\rho^2>\mu^2), \nn\\
F(\rho) &=& \frac{\rho}{\mu}{\rm artanh}\frac{\rho}{\mu} \quad (\rho^2<\mu^2),
 \ea
both being singular for $\rho^2=\mu^2$. In both cases, the solution (\ref{XaFbrho}) is subject to the constraint
(\ref{R21}), which is solved by the constraints
 \be\lb{a0bs1}
\a^2 = \a\cdot\b = 0, \quad \b^2 = \frac4{l^2}.
 \ee
The null vector $\J$ is $\J=\a\wedge\b=\varepsilon\a$.

The AdS wave solution given in equations
(4) and (9) of \cite{GT} corresponds to the first solution (\ref{Feta-}), with $\eta<0$ and $\rho \propto r^{-2}$.
For $\eta>0$, (\ref{Feta+}) leads to a rotating black hole solution. Let us parameterize the vectors $\a$ and $\b$ subject
to the the constraints (\ref{a0bs1}) by
\[
\a = \frac{M}{4}(1+l^2,1-l^2,\pm 2l), \quad \b = (1-l^{-2},-1-l^{-2},0).
\]
This choice is motivated by the observation that for $C\to0$ ($\lambda\to0$),
corresponding to a vanishing scalar field, the equations reduce to those of vacuum AdS gravity,
with the solution (\ref{XaFbrho}) where $F(\rho)\to1$ and $\a$ and $\b$ obey the constraints
(\ref{a0bs1}), corresponding to an extreme BTZ black hole \cite{GC94b} of mass $M$ and angular momentum
$J=\mp Ml$. So the solution can be viewed as an extreme BTZ black hole dressed by a Galileon scalar field.
Putting $\rho=r^2/2$ and $MF(\rho)=M(r)$, the solution takes the form
  \ba\lb{arctan}
ds^2 &=& - \frac{r^4dt^2}{l^2(r^2+l^2M(r)/2)} + \frac{l^2dr^2}{r^2} + (r^2+l^2M(r)/2)
\left(d\varphi \pm \frac{lM(r)/2}{r^2+l^2M(r)/2}dt\right)^2, \nn\\
\phi(r) &=& -\frac{Cl^4}{4\eta r^2}, \qquad M(r) = \frac{lM}{\eta^{1/2}\phi(r)}
\arctan\left(\frac{\eta^{1/2}\phi(r)}{l}\right).
 \ea

For $M=0$ the metric reduces to that of the BTZ vacuum. For $M>0$, $M(r)$ is positive, so that
the metric (\ref{arctan}) is everywhere causal ($g_{\varphi\varphi}>0$), except
for $r=0$, corresponding to a degenerate horizon. The near-horizon metric is
 \be
ds^2 \simeq - \frac{r^2dt^2}{l^2+b^2} + \frac{l^2dr^2}{r^2} + \frac{(l^2+b^2)r^2}{l^2}\left(d\varphi \pm
\frac{b^2}{l(l^2+b^2)}dt\right)^2
 \ee
(with $b^2= \pi\eta^{1/2}lM/|C|$), consistent with the generic behavior (\ref{nh}). The asymptotic
behavior is that of the extreme BTZ metric (transform $r^2$ to $r^2 = \overline{r}^2 - l^2M/2$).
It follows from the constraints (\ref{a0bs1}) that no scalar invariant constructed from the metric
can depend on $F(\rho)$, so that the curvature invariants
 \be
\R = -6l^{-2}, \qquad \R^{\mu\nu}\R_{\mu\nu} = 12l^{-4}
 \ee
are those of extreme BTZ or of $AdS_3$, which suggests that the spacetime may be singularity free.
This conclusion may be strengthened by studying the first integrated geodesic equation \cite{GC94b}
 \be
R^{-2}\dot\rho^2 + \lambda^{ab}\Pi_a\Pi_b = \varepsilon,
 \ee
where the dot denotes derivative with respect to an affine parameter $\tau$, $\Pi_0=E$ and $\Pi_1=L$
are constants of the motion, and $\varepsilon= -1$, $0$ or $+1$ for timelike, null or spacelike geodesics.
For the metric (\ref{arctan}) this takes the form
 \be
\dot{r}^2 + \frac{L^2}{l^2} - E^2 - \varepsilon\,\frac{r^2}{l^2} = \left(E\mp\frac{L}{l}\right)^2B(r),
 \ee
where
 \be
B(r) = \frac{2\eta^{1/2}M}{Cl}\,\arctan\left(\frac{Cl^3}{4\eta^{1/2}r^2}\right)
 \ee
is everywhere bounded. It follows that the maximally extended spacetime ($r$ real) is geodesically complete.\\

\noindent b) \underline{$\X'^2+12\Lambda=\pm(-12/\eta)^{1/2}C/R$} \\

We can treat together these two possibilities by putting, as in subsection 4.2, $\gamma^2 = -3C^2/\eta$
($\gamma$ real). We then obtain
 \be\lb{51b}
W = 6\Lambda - \frac\gamma{R}, \quad \Psi = \frac23, \quad \Omega = - \frac{2\gamma R}3
 \ee
(the value of $\Psi$ is easily obtained by substituting $\eta\Lambda=-\alpha/3$ in (\ref{psi1})).
Because $\Psi$ is constant, equation (\ref{J}) reduces to the conservation of the ``orbital''
angular momentum $\L = \X\wedge\X' = 3\J/2$. The vector field equation (\ref{ein1}) and the associated
first integral (the Hamiltonian constraint) also become extremely simple:
 \be
\X'' = -\gamma\frac\X{R^3}, \qquad \frac12\X'^2 - \frac\gamma{R} = -6\Lambda.
 \ee
We recognize the equations of motion of a massive particle in a central Newtonian field (the Kepler problem),
with the difference that the target 3-space metric is Lorentzian. However the present problem can be treated
by the same methods if \underline{$\L^2\neq0$}.

As mentioned in Sect. 4, there is no horizon for $\L^2<0$, so we assume $\L^2>0$ ($\L$ spacelike). Then we
can rotate the axes so that only the component $L_Y$ is non-zero and the motion is confined to the plane
($T,X$), i.e. the physical metric is static. This can be parameterized by (\ref{para}) with
 \be
\X = (R\sinh\chi, -R\cosh\chi, 0).
 \ee
The associated constant orbital angular momentum is
 \be
R^2\chi'= L.
 \ee
It follows that $R'= L\dot{R}/R^2$, where the dot indicates a derivative with respect to $\chi$. Transforming
to the radial coordinate $\chi$, the metric reads
 \be\lb{kepler}
ds^2 = - Re^{-\chi}dt^2 + Re^{\chi}d\varphi^2 + L^{-2}R^2d\chi^2,
 \ee
with $R(\chi) = u^{-1}(\chi)$, where $u(\chi)$ obeys the equation, derived from the radial equation
(\ref{radial}):
 \be
\dot{u}^2 = u^2 + \frac{2\gamma}{L^2}u - \frac{12\Lambda}{L ^2}.
 \ee
This is solved by
 \be\lb{solkepler}
R(\chi) = \left[bF(\chi) - \gamma/L^2\right]^{-1},
 \ee
with
 \be\lb{Fkepler}
F = \cosh\chi \;\; (\Delta>0), \quad F = e^\chi \;\; (\Delta=0), \quad F = \sinh\chi \;\; (\Delta<0),
 \ee
where $\Delta = L^{-4}(\gamma^2+12\Lambda L^2)$.

If \underline{$\L^2=0$}, we know that $\X$ is orthogonal to $\L=L\a$, so that
 \be\lb{kepler1}
\X = \a F + \b R, \qquad \a^2 = \a\cdot\b = 0,
 \ee
where
 \be
R'^2 = \frac{2\gamma-12\Lambda R}R, \quad F'' + \frac\gamma{R^3}F = 0.
 \ee
There can be a Killing horizon if $\gamma>0$, however both this solution and solution (\ref{kepler}) are such
that $\X'^2 \propto R^{-1}$ near the horizon, and so, according to the analysis of subsection 4.2, are
singular on the horizon.

\subsection{Case $\alpha=0$}
In this case, equation (\ref{ham1}) factorizes as
 \be
\X'^2\left[\X'^2(\X'^2+4\Lambda) + \frac{12C^2}{\eta R^2}\right] = 0.
 \ee

But $\X'^2=\alpha=0$ is excluded because we have assumed $C\neq 0$ (recall eq. (\ref{phi})). So there
remains the possibility
 \be\lb{a0}
W = \Lambda\left[1\pm\sqrt{1-\frac{3C^2}{\eta\Lambda^2R^2}}\right], \quad
\Psi =\frac23\left(1+\frac\Lambda{W}\right), \quad \Omega = -\frac{2C^2}{\eta W},
 \ee
where we have assumed $\Lambda\neq0$. For $\eta<0$ this will lead to two solutions
with singular horizons, as already discussed. For $\eta>0$, on the other hand,
$R$ cannot vanish, but must have a minimum value $R_0\ge\delta$ (where we have put
$\delta^2=3C^2/\eta\Lambda^2$) such that
 \be\lb{sol0}
R'^2(R_0) = \frac{J^2}{R_0^2\Psi^2(R_0)} - 2W(R_0) = 0.
 \ee
If $\Lambda<0$, we show in the Appendix that, for any $R_0>\delta$ there are two negative values
of $J^2$ satisfying this equation and such that $dR'^2/dR(R_0) > 0$, corresponding to two soliton solutions
with different asymptotic behaviors.

The subcase $\alpha=\Lambda=0$ is also a subcase of the case $\Lambda+\alpha/3\eta=0$ (subcase b) with $\Lambda=0$), the expressions
of the functions $W$, $\Psi$ and $\Omega$ being given by (\ref{51b}) ($\eta<0$) for $\Lambda=0$. For $J^2>0$ the
solution is given by (\ref{kepler}), (\ref{solkepler}) and (\ref{Fkepler}) with $\Delta>0$. For $J^2=0$ the exact
solution is given by (\ref{Winf2}), both solutions being again singular on the horizon.

\section{Conclusion}
We have investigated the whole range of stationary rotationally symmetric solutions to the truncated
Horndeski action (\ref{ac}) in three spacetime dimensions. The special case of a vanishing scalar charge with
non-vanishing minimal scalar coupling constant (which includes the BTZ black hole solution) was previously discussed in
\cite{BGH2}. We have extended here the discussion of the degenerate sector to the non-static case. We have also
found a new degenerate sector corresponding to the case where the minimal scalar coupling constant and the scalar
charge both vanish.

In the physically more interesting case where the scalar charge is non-zero, we have investigated the near-horizon
behavior and the asymptotic behavior of the solutions for generic values of the model parameters. While there may be
in principle up to three different solutions with Killing horizon for the same values of the model coupling constants and of
the scalar charge, only one of these can correspond to a black hole with scalar hair
(provided the two scalar coupling constants have the same sign), with the near-horizon behavior (\ref{nh}),
a constant curvature asymptotic behavior
being consistent with all the field equations. The horizon is always degenerate. In the special case where the
three model parameters are constrained by $\Lambda+\alpha/3\eta=0$, $\alpha>0$, $\eta>0$, we have been able to find
analytically a new rotating black hole solution (\ref{arctan}) with degenerate horizon. This solution, which
depends on two parameters (mass and scalar charge), is interpreted as an extreme BTZ black hole dressed by a Galileon
scalar field. It is geodesically and causally complete, and asymptotic to the
extreme BTZ metric. The scalar field is singular on the horizon. A by-product of our analysis is the existence of
soliton solutions, discussed briefly in the Appendix in the case of a vanishing minimal scalar coupling constant.

It would be interesting to extend this work to the search for charged Galileon black holes \cite{bhemgal} in three
dimensions, following the framework developed in \cite{GC93}.

\section*{Acknowledgments}
One of us (K.N) thanks the LAPTH Annecy for warm hospitality and the University of Jijel for financial support.

\renewcommand{\theequation}{A.\arabic{equation}}
\setcounter{equation}{0}
\section*{Appendix: Soliton solutions (Case $\alpha=0$)}
For $\alpha=0$, $\eta>0$, the effective potential $W(R)$ is of the form
 \be\lb{Wsol}
W_\pm = \Lambda x, \quad x \equiv 1\pm\sqrt{1-\delta^2/R^2} \quad (0<x<2),
 \ee
with $\delta^2=3C^2/\eta\Lambda^2$. Using this and (\ref{a0}), we find from the radial equation
(\ref{radial}) that
 \be\lb{y(x)}
R'^2 = \frac{9J^2}{4\delta^2} y, \quad y(x) \equiv \frac{x^3(2-x)}{(x+1)^2} - \lambda^2x \quad
\left(\lambda^2 = \frac{8\Lambda\delta^2}{9J^2}\right).
 \ee
For a given real value of $J^2$, the minimum value $R_0$ of $R$ is such that $R'^2(R_0) =0$ ($y(x_0)=0$)
and $2R''(R_0)=dR'^2/dR(R_0) > 0$ (sign($dy/dx(x_0)) = \pm{\rm sign}(J^2)$). The derivative of (\ref{y(x)}) is
 \be
\frac{dy}{dx} = \frac{2x^2(-x^2-x+3)}{(x+1)^3} - \lambda^2.
 \ee
Replacing $\lambda^2$ with its value from  $y(x_0)=0$,
 \be
\lambda^2(x_0) = \frac{x_0^2(2-x_0)}{(x_0+1)^2}
 \ee
(which must be positive, implying $\Lambda J^2>0$), we obtain
 \be
\frac{dy}{dx}(x_0) = -\frac{x_0^2(x_0-1)(x_0+4)}{(x_0+1)^3},
 \ee
which is of the sign of $-(x_0-1)$, i.e. $\mp$. So, for both branches, the condition for $R_0$ to correspond to a
minimum is $J^2<0$, implying $\Lambda<0$.

The curve $\lambda^2(x_0)$ is convex for $0<x_0<2$, with a maximum at $x_0=1$. So, for a given value of $J^2$
(of $\lambda^2$), there are two values of $x_0$, one on the branch $(+)$, the other on the branch $(-)$. The
asymptotic behaviors of the corresponding solutions are different. On the $(+)$ branch, $R\to\infty$ corresponds
to $x\to2$, leading to $\X'^2 \sim R'^2 \to -4\Lambda$. Together with $\Psi(\infty)=1$, this means that the
resulting metric is asymptotically AdS. On the $(-)$ branch, (\ref{Wsol}) together with (\ref{radial}) lead to
$\X'^2 \sim R'^2 \sim -\Lambda\delta^2/R^2$, so that $R^2 \sim a\rho$ ($a^2=-4\Lambda\delta^2$). The integration
of (\ref{ein1}) with $\Psi \sim (4a/3\delta^2)\rho$, $\Omega \sim (a^3/3\delta^2)\rho$ then leads to
 \be
\X \sim \a\rho^{1/2} + \b\rho^{-1/2} \qquad \left(\a^2 = a^2\right).
 \ee
Putting $\rho=ar^2/4$, the asymptotic metric is, up to a local coordinate transformation,
 \be
ds^2 \sim -brdt^2 + dr^2 + crd\varphi^2
 \ee
($b$ and $c$ positive constants).

\end{document}